\documentclass[12pt,english,a4]{article}
\usepackage{mathptmx}

\usepackage[T1]{fontenc}
\usepackage[utf8]{inputenc}
\usepackage{babel}
\usepackage{amsmath}
\usepackage{amssymb}
\usepackage[unicode=true,pdfusetitle,
 bookmarks=true,bookmarksnumbered=false,bookmarksopen=false,
 breaklinks=false,pdfborder={0 0 1},backref=false,colorlinks=false]
 {hyperref}

\makeatletter
\usepackage{babel}

\makeatother

\begin{document}

\title{Partial Transpose as a General Criterion for the Separability of
Quantum States}

\author{{\footnotesize{{PRANAV.P , M.RAVENDRANADHAN}}}%
\thanks{{\footnotesize{ravicusat@cusat.ac.in, Please make comments on this
paper}}%
}\\
 {\footnotesize{{DEPARTMENT OF PHYSICS}}}\\
 {\footnotesize{{COCHIN UNIVERSITY OF SCIENCE AND TECHNOLOGY}}}\\
 {\footnotesize{{KOCHI, KERALA, INDIA}}}\\
 {\footnotesize{{PIN-682002}}}}
\maketitle
\begin{abstract}
Usual separability criteria applicable to distinguishable particles
are not applicable to identical particles. In this paper, we will
show that partial transpose operation and symmetrization (anti symmetrization)
of density matrix of bipartite boson system (bipartite fermion system)
give result which can be used to check whether a state is separable
or not. By using determinant based separability criteria it has been
found that for identical particles, whatever be the Schmidt's number
(for bosons) or Slater rank (for fermions) the state is separable.
It is found that partial transposition and symmetrization (antisymmetric)
is equivalent to the matrix realignment method proposed by Wu. We
will show that this separability criterion can also be applied to
distinguishable particles. 
\end{abstract}

\subsection*{Introduction}

The property of quantum entanglement or the inseparability of quantum
states lies at the heart of several fields like quantum information
theory, quantum teleportation, quantum cryptography etc. However for
a given a quantum state it is a non trivial question to find out whether
a given state is entangled or not. So several criterion's such as
Peres Horodecki Positivity of Partial Transpose (PPT) criterion\cite{peres 2},
Von Neumann entropy, Schmidt decomposition etc. have already been
proposed to find out whether a given state is entangled or not. However,
the entanglement of indistinguishable particles are not well studied
since wave function of such particles are symmetrised or antisymmetrised
product states which may not be separable in the usual sense. But
these inseparability does not imply entanglement as it need not lead
to any useful correlations. So the above mentioned separability criterion
may not work in the case of indistinguishable particles. Von Neumann
entropy\cite{ghirardi-2} is one generalized criterion which works
for distinguishable particles. But the entropy does change for indistinguishable
particles. Slater decomposition or slater rank is another one criterion
proposed for fermions. Another test for separability is the Positivity
of Partial transpose (PPT) proposed by Peres and Horodoki\cite{peres 2,Horodoki}.
According to this criteria, for the separability, partially transposed
density matrix should be positive. Another simple criteria, similar
to this is proposed by Augusiak \emph{et al }\cite{Augusiak} based
on the determinant of the partially transposed density matrix. According
to this criteria, for the separability, determinant of the partially
transposed density matrix should be positive.

Recently a separability criteria was proposed by Zhao \emph{et al}(\cite{Xin})
known as \emph{Partial Transposed Hermitian conjugation criterion.
}They proposed that for the separability the density matrix should
be invariant under the operation of partial transposed conjugation
operation on a member in the bipartite system. Their proposition can
be obtained as follows. If a bipartite system consisting of particle
$A$ and particle $B$ is separable, density matrix can be written
as 
\[
\rho=\rho_{A}\otimes\rho_{B}
\]
But we know that for any particle $\rho_{A,B}=\rho_{A,B}^{\dagger}$.
Therefore in the above equation we can write $\rho_{B}=\rho_{B}^{\dagger}$.
Then 
\[
\rho=\rho_{A}\otimes\rho_{B}^{\dagger}
\]
That is we take the transposed conjugation of the system $B$ only
(partial Hermitian conjugation). Therefore for the separable system
\[
\rho=\rho^{PHC}
\]

In this paper, we will show that partial transpose operation and symmetrization
(antisymmetrization) of density matrix of bipartite boson system (bipartite
fermion system) give result which can be used to check whether a state
is separable or not. Here we used determinant based separability test
prosed by Augusiak\emph{ et al}\cite{Augusiak} to check the positivity
of partially transposed density matrix. Our result shows that if the
particles are identical in all sense, the states are separable.

It is found that partial transposition and symmetrization (antisymmetric)
is equivalent to the matrix realignment proposed by Wu. We will show
that this separability criterion can also be applied to distinguishable
particles.

\subsection*{Indistinguishable Particles}

Hilbert space of two distinguishable particles, $\mathcal{H=H_{\textnormal{1}}\otimes H}_{2}$.
Where $\mathcal{H}_{\textnormal{1}}$ belongs to particle one and
$\mathcal{H}_{2}$ belongs to particles two. For distinguishable particles
entanglement is attributed to states which cannot be written in product
form. But a system of indistinguishable particle in general, cannot
be written in product form because both the particles share the same
Hilbert space. Therefore many of the criteria that is discussed for
distinguishable particles is not applicable. For Bosons and Fermions
inseparable states with VonNeuman entropy equal to one can be shown
to be separable. But this inseparability does not mean entanglement.
So special care is to be taken when entanglement of indistinguishable
particles are studied.

\subsubsection*{Bosons}

A bipartite boson system in the Schmidt basis can be written as 
\begin{equation}
\left|\psi\right\rangle =\sum_{i=1}^{n}d_{i}a_{i}^{\dagger}a_{i}^{\dagger}\left|0\right\rangle \label{eq:1a}
\end{equation}

Where $d_{i}$ is real and number of nonzero $d_{i}$ is the Schmidt
number. Equation (\ref{eq:1a}) represents a state with Schimidts
number $n.$ Then density matrix 
\begin{equation}
\rho=\sum_{i,j=1}^{n}d_{i}d_{j}a_{i}^{\dagger}a_{i}^{\dagger}\left|0\right\rangle \left|0\right\rangle \left\langle 0\right|a_{j}a_{j}\label{eq:14}
\end{equation}
\begin{equation}
\rho^{PT}=\sum_{i,j=1}^{n}d_{i}d_{j}a_{i}^{\dagger}a_{j}^{\dagger}\left|0\right\rangle \left|0\right\rangle \left\langle 0\right|a_{i}a_{j}\label{eq:14a}
\end{equation}
In equation (\ref{eq:1a}) $d_{i}$ is real and then $\rho^{PT}=\rho^{PHC}.$
For Schmidt's number equal to one, 
\[
\rho=\rho^{PT}
\]
and then the state is separable as expected. For Schmidt's number
greater than 1, from (\ref{eq:14}) and (\ref{eq:14a}) $\rho\neq\rho^{PT}$
and then the state in (\ref{eq:1a}) appear to be entangled. We can
show that this state is separable even when Schmidt's number is $\geq2.$

For bosons for $i\neq j$ 
\begin{equation}
\left[a_{i}^{\dagger},a_{j}^{\dagger}\right]=0\label{eq:16}
\end{equation}
and then after symmetrization (that is by exchanging $i$ and $j$
in the last part of equation (\ref{eq:14a}) we get 
\[
\rho^{PT}=\sum_{ij}d_{i}d_{j}a_{i}^{\dagger}a_{j}^{\dagger}\left|0\right\rangle \left|0\right\rangle \left\langle 0\right|a_{j}a_{i}=\sum_{i}d_{i}a_{i}^{\dagger}\left(\sum_{j}d_{j}a_{j}^{\dagger}\left|0\right\rangle \left|0\right\rangle \left\langle 0\right|a_{j}\right)a_{i}
\]
\[
=\rho_{1}\otimes\rho_{2}
\]
To check the positivity, we take determinant on both sides and get
\[
\det\rho^{PT}=\left(\det\rho_{1}\right)^{n}\left(\det\rho_{2}\right)^{n}=\left(d_{1}d_{2}\cdots\cdots d_{n}\right)^{2n}>0
\]

This result guarantees positivity of partial transpose \cite{peres 2,Horodoki}
. Then we may conclude that bipartite bosonic states are separable
for all values of Schmidt number. Ghirardi \cite{ghirardi-2} had
shown that states are separable when the Schmidt number is 1 or 2
but he hadn't shown that it is entangled when the Schmidt number is
greater than 2. That is bipartite bosonic states are always separable.
For getting entanglement there should be some quantum numbers which
makes the particles distinguishable. For example the state 
\[
\left|\psi\right\rangle =\frac{\left|\uparrow\downarrow\right\rangle \pm\left|\downarrow\uparrow\right\rangle }{\sqrt{2}}
\]
is separable. But the state
\[
\left|\psi\right\rangle =\frac{\left|L\uparrow,R\downarrow\right\rangle \pm\left|L\downarrow,R\uparrow\right\rangle }{\sqrt{2}}
\]
is entangled. Where $L$ and $R$ are space indices's, left and right. 

This issue is discussed in \cite{cirac-1}. As an example let us consider
the case with Schmidt's number 2, 
\[
\left|\psi\right\rangle =d_{i}a_{i}^{\dagger}a_{i}^{\dagger}\left|0\right\rangle +d_{j}a_{j}^{\dagger}a_{j}^{\dagger}\left|0\right\rangle 
\]
\[
\rho=\left(d_{i}a_{i}^{\dagger}a_{i}^{\dagger}\left|0\right\rangle +d_{j}a_{j}^{\dagger}a_{j}^{\dagger}\left|0\right\rangle \right)\left(\left\langle 0\right|a_{i}a_{i}\ d_{i}+\left\langle 0\right|a_{j}a_{j}\ d_{j}\right)
\]
\begin{equation}
=\begin{alignedat}{1}d_{i}^{2}a_{i}^{\dagger}a_{i}^{\dagger}\left|0\right\rangle \left\langle 0\right|a_{i}a_{i} & +d_{j}^{2}a_{j}^{\dagger}a_{j}^{\dagger}\left|0\right\rangle \left|0\right\rangle \left\langle 0\right|a_{j}a_{j}\\
+\\
d_{i}d_{j}a_{i}^{\dagger}a_{i}^{\dagger}\left|0\right\rangle \left|0\right\rangle \left\langle 0\right|a_{j}a_{j} & +d_{i}d_{j}a_{j}^{\dagger}a_{j}^{\dagger}\left|0\right\rangle \left|0\right\rangle \left\langle 0\right|a_{i}a_{i}
\end{alignedat}
\label{eq:15}
\end{equation}
\begin{equation}
=\left(\begin{array}{cccc}
d_{i}^{2} & 0 & 0 & d_{i}d_{j}\\
0 & 0 & 0 & 0\\
0 & 0 & 0 & 0\\
d_{i}d_{j} & 0 & 0 & d_{j}^{2}
\end{array}\right)\label{eq:15a}
\end{equation}
In this case $\det\rho=0.$ But if we take 
\begin{equation}
\rho^{PT}=\begin{alignedat}{1}\{d_{i}^{2}a_{i}^{\dagger}a_{i}^{\dagger}\left|0\right\rangle \left\langle 0\right|a_{i}a_{i} & +d_{j}^{2}a_{j}^{\dagger}a_{j}^{\dagger}\left|0\right\rangle \left|0\right\rangle \left\langle 0\right|a_{j}a_{j}\\
+\\
d_{i}d_{j}a_{i}^{\dagger}a_{j}^{\dagger}\left|0\right\rangle \left|0\right\rangle \left\langle 0\right|a_{i}a_{j} & +d_{i}d_{j}a_{j}^{\dagger}a_{i}^{\dagger}\left|0\right\rangle \left|0\right\rangle \left\langle 0\right|a_{j}a_{i}\}
\end{alignedat}
\label{eq:18}
\end{equation}
\[
=\left(\begin{array}{cccc}
d_{i}^{2} & 0 & 0 & 0\\
0 & 0 & d_{i}d_{j} & 0\\
0 & d_{i}d_{j} & 0 & 0\\
0 & 0 & 0 & d_{j}^{2}
\end{array}\right)
\]
and $\det\rho^{PT}=-d_{i}^{4}d_{j}^{4}.$ Then the state may appear
to be entangled.

By using (\ref{eq:16}) (that is symmetrization) we can write, 
\[
\rho^{PT}=\begin{alignedat}{1}\{d_{i}^{2}a_{i}^{\dagger}a_{i}^{\dagger}\left|0\right\rangle \left\langle 0\right|a_{i}a_{i} & +d_{j}^{2}a_{j}^{\dagger}a_{j}^{\dagger}\left|0\right\rangle \left|0\right\rangle \left\langle 0\right|a_{j}a_{j}\\
+\\
d_{i}d_{j}a_{i}^{\dagger}a_{j}^{\dagger}\left|0\right\rangle \left|0\right\rangle \left\langle 0\right|a_{j}a_{i} & +d_{i}d_{j}a_{j}^{\dagger}a_{i}^{\dagger}\left|0\right\rangle \left|0\right\rangle \left\langle 0\right|a_{i}a_{j}\}
\end{alignedat}
\]
\begin{equation}
=\left(\begin{array}{cccc}
d_{i}^{2} & 0 & 0 & 0\\
0 & d_{i}d_{j} & 0 & 0\\
0 & 0 & d_{i}d_{j} & 0\\
0 & 0 & 0 & d_{j}^{2}
\end{array}\right)=\left(\begin{array}{cc}
d_{i} & 0\\
0 & d_{j}
\end{array}\right)\otimes\left(\begin{array}{cc}
d_{i} & 0\\
0 & d_{j}
\end{array}\right)=\rho_{1}\otimes\rho_{2}\label{eq:17}
\end{equation}
Evidently $\det\rho^{PT}=d_{i}^{4}d_{j}^{4}>0$ and the state is separable.
 That is it is possible to write it as product of density matrices.
This will guarantee the positivity of $\rho^{PT}$ and then the state
is separable\cite{ghirardi-2}.

Equation (\ref{eq:17}) can be obtained from equation (\ref{eq:15a})
by matrix realignment method introduced by Wu and Yang\cite{Wu}.
That is partial transposition and symmetrization can be considered
equivalent to realignment of density matrix.

\subsubsection*{Fermions}

A generic state for a two fermion system can be represented as 
\begin{equation}
\left|\psi\right\rangle =\sum_{ij}\Omega_{i,j}f_{i}^{\dagger}f_{j}^{\dagger}\left|0\right\rangle \label{eq:8}
\end{equation}

where $f^{\dagger}$ is the fermionic creation operating on the vacuum
state $\left|0\right\rangle $ to create the fermions and $\Omega_{ij}=-\Omega_{ji}$
is an antisymmetric matrix. Similar to the Schmidt decomposition used
before, there exist a decomposition\cite{mehta,schliemann} for antisymmetric
matrices known as slater decomposition by which the above state can
be written as 
\begin{equation}
\left|\psi\right\rangle =\sum_{l=1}^{n}z_{l}f_{1_{l}}^{\dagger}f_{2_{l}}^{\dagger}\left|0\right\rangle \label{eq:9}
\end{equation}
Where the number of non vanishing coefficients $z_{l}$ gives the
Slater rank. Here $z_{l}$ may not be real. For slater rank 1, the
density matrix is given by 
\begin{equation}
\rho=z^{2}f_{1_{i}}^{\dagger}f_{2_{i}}^{\dagger}\left|0\right\rangle \left\langle 0\right|f_{2_{i}}f_{1_{i}}=\rho^{PT}=\rho_{1}\otimes\rho_{2}..\label{eq:10}
\end{equation}
The density matrix has only one element and hence it corresponds to
a separable state which is expected for a system with slater rank
1. In general, for the state with Slater rank $n$ in equation (\ref{eq:9})
\[
\rho=\sum_{i,j=1}^{n}z_{i}z_{j}^{*}f_{1_{l}}^{\dagger}f_{2_{l}}^{\dagger}\left|0\right\rangle \left\langle 0\right|f_{2_{j}}f_{1_{j}}
\]
Then 
\[
\rho^{PT}=\sum_{i,j}z_{i}z_{j}^{*}f_{1_{l}}^{\dagger}f_{2_{j}}^{\dagger}\left|0\right\rangle \left\langle 0\right|f_{2_{i}}f_{1_{j}}
\]
It can be written as a matrix
\[
\rho^{PT}=\left(\begin{array}{cccccccccc}
\left|z_{1}\right|^{2} & 0 & 0 & 0 & 0 & 0 & 0 & 0 &  & 0\\
0 & -z_{1}z_{2}^{*} & 0 & 0 & 0 & 0 & 0 & 0 &  & 0\\
0 & 0 & -z_{2}z_{1}^{*} & 0 & 0 & 0 & 0 & 0 &  & 0\\
0 & 0 & 0 & \left|z_{2}\right|^{2} & 0 & 0 & 0 & 0 &  & 0\\
0 & 0 & 0 & 0 & -z_{2}z_{3}^{*} & 0 & 0 & 0 &  & 0\\
0 & 0 & 0 & 0 & 0 & -z_{3}z_{2}^{*} & 0 & 0 &  & 0\\
0 & 0 & 0 & 0 & 0 & 0 & \left|z_{3}\right|^{2} & 0 & \cdots & 0\\
0 & 0 & 0 & 0 & 0 & 0 & 0 & \ddots & \ldots & 0\\
\vdots & \vdots &  &  &  &  &  & \vdots & \ddots & \vdots\\
0 & 0 & 0 & 0 & 0 & 0 & 0 & 0 &  & \left|z_{n}\right|^{2}
\end{array}\right)
\]
Here we used $\left\{ f_{i}^{\dagger},f_{j}^{\dagger}\right\} =0,$
for fermions. When $i=j$, no anti symmetrization is needed and that
is why there is no negative sign with $\left|z_{i}\right|^{2}$ terms.
Evidently
\begin{equation}
\det\rho^{PT}=\left|z_{1}\right|^{4}\left|z_{2}\right|^{4}\cdots\cdots\left|z_{n}\right|^{4}\label{eq:13}
\end{equation}
$\det\rho^{PT}>0$ and the two fermion state with Slater rank $n$
is entangled.

\subsection*{Distinguishable Particles}

In Schmidt's basis a bipartite quantum state can be represented as
\begin{equation}
\left|\psi\right\rangle =\sum_{i}\omega_{i}a_{i}^{\dagger}b_{i}^{\dagger}\left|0\right\rangle .\label{eq:2}
\end{equation}

Where $\omega_{i}$ is real. Density matrix for the system in the
Schmidt basis is

\begin{equation}
\rho=\sum_{ij}\omega_{i}\omega_{j}a_{i}^{\dagger}b_{i}^{\dagger}\left|0\right\rangle \left\langle 0\right|b_{j}a_{j}\label{eq:3}
\end{equation}

Then 
\begin{equation}
\rho^{PT}=\sum_{ij}\omega_{i}\omega_{j}a_{i}^{\dagger}b_{j}^{\dagger}\left|0\right\rangle \left\langle 0\right|b_{i}a_{j}\label{eq:4}
\end{equation}
For Schmidt's number one, 
\[
\rho=\rho^{PT}
\]
and then the state is separable.

In general $\rho\neq\rho^{PT}$ and hence the state is not separable.

\subsubsection*{Conclusion}

In summary, we have presented a general criteria for the separability
of quantum states for a bipartite system based on partial transposition
operation for both distinguishable and indistinguishable particles.
For bosons, the partial transpose is taken on the Schmidt basis, while
it is done using slater decomposition for fermions. It has been found
that for identical particles, whatever be the Schmidt's number (for
bosons) or Slater rank (for fermions) the state is separable. It is
found that partial transposition and symmetrization (antisymmetrization)
is equivalent to the matrix realignment method proposed by Wu. We
will show that this separability criterion can also be applied to
distinguishable particles.

\end{document}